\pdfoutput=1
\documentclass[11pt]{article}

\usepackage{ACL2023}
\usepackage{booktabs}
\usepackage{times}
\usepackage{latexsym}
\usepackage{array,multirow}
\usepackage[T1]{fontenc}
\usepackage[utf8]{inputenc}

\usepackage{microtype}

\usepackage{inconsolata}

\usepackage{graphicx}
\usepackage{dblfloatfix}
\usepackage{enumitem}
\usepackage{float}
\usepackage{amsmath,amssymb}

\DeclareMathOperator*{\argmin}{argmin}

\title{Query Encoder Distillation via Embedding Alignment is a Strong Baseline Method to Boost Dense Retriever Online Efficiency}
\author{
  Yuxuan Wang \and Hong Lyu\thanks{\ \ Both authors contributed equally to this research.} \\
  University of Pennsylvania \\
  \texttt{\{wangy49, hlyu\}@seas.upenn.edu}
}

\begin{document}
\maketitle
\begin{abstract}
The information retrieval community has made significant progress in improving the efficiency of Dual Encoder (DE) dense passage retrieval systems, making them suitable for latency-sensitive settings. However, many proposed procedures are often too complex or resource-intensive, which makes it difficult for practitioners to adopt them or identify sources of empirical gains. Therefore, in this work, we propose a trivially simple recipe to serve as a baseline method for boosting the efficiency of DE retrievers leveraging an asymmetric architecture. Our results demonstrate that even a 2-layer, BERT-based query encoder can still retain 92.5\% of the full DE performance on the BEIR benchmark via unsupervised distillation and \textbf{proper student initialization}. We hope that our findings will encourage the community to re-evaluate the trade-offs between method complexity and performance improvements.
\end{abstract}
\section{Introduction}
  % Information Retrieval (IR) is a crucial component of various Natural Language Processing (NLP) systems, such as question-answering \citep{chen-etal-2017-reading, DBLP:journals/corr/abs-2004-04906, DBLP:journals/corr/abs-2112-09332} and chatbots \citep{shuster2022blenderbot}. 
  
  Recent advances in neural-based NLP techniques have led to powerful neural encoders that can generate high-quality, semantic-rich, dense vector text representations \citep{reimers_sentence-bert_2019, cer_universal_2018, conneau_supervised_2018, schick2023toolformer}, making it possible to calculate the text relevancy with simple vector operations like dot product. Thus, the Dual Encoder (DE) neural Information Retrieval (IR) architectures, combined with optimized semantic search implementations \citep{DBLP:journals/corr/abs-1806-09823, faiss, DBLP:journals/corr/abs-2010-14848}, have achieved comparable or even superior performances to their Cross Encoder (CE) based predecessors \citep{thakur_beir_2021, menon_defense_2022, ni-etal-2022-large, yu2022coco} while being significantly more efficient \citep{reimers_sentence-bert_2019}.

  Despite the numerous proposed efficiency enhancements for making DE-based IR models suitable for production settings, they may pose challenges for practitioners with limited resources in terms of adoption and replication \citep{hardware-lottery}. However, by leveraging two key facts, we can simplify model development while achieving higher efficiency. Firstly, documents, in contrast to queries, are typically longer and more complex, necessitating specialized architectures \citep{DBLP:journals/corr/abs-1912-08777, DBLP:journals/corr/abs-1901-02860, DBLP:journals/corr/abs-2004-05150, DBLP:journals/corr/abs-2007-14062}. Secondly, document embeddings remain mostly static after indexing, allowing for a high-quality and computationally expensive document encoder without online overhead. Based on these insights, we propose an asymmetric IR architecture that pairs a lightweight query encoder with a robust document encoder.
  
\begin{figure*}[htbp!]
    \centering
    \includegraphics[width=\textwidth]{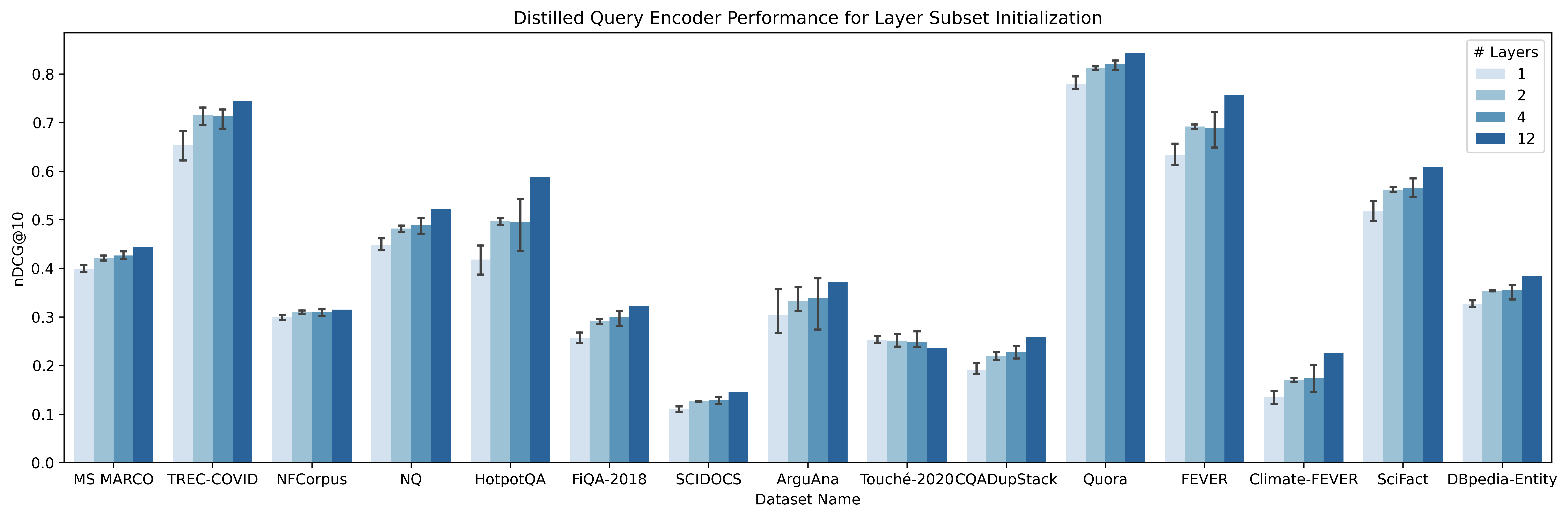}
    \caption{The dual encoder retriever performance with distilled query encoders of a varying number of layers. The student models are initialized by extracting subsets of the teacher model (\texttt{\textcolor{blue}{msmarco-bert-base-dot-v5}}) layers. Variances in performances come from different layer subsets chosen, discussed in \autoref{sec:results}.}
    \label{fig:extractive-performance}
  \end{figure*}

  In this study, we present a minimalistic baseline approach for constructing the aforementioned asymmetric retriever using any existing query encoder. As depicted in \autoref{fig:extractive-performance}, by employing suitable initialization and simply minimizing the Euclidean distance between student and teacher query embeddings, even a 2-layer BERT-based query encoder \citep{bert} can retain 92.5\% of the full DE performance on the BEIR benchmark \citep{thakur_beir_2021}. Similarly, the 4-layer encoder preserves 96.2\% of the full performance, which aligns with the supervised outcome (96.6\%) achieved by a 6-layer encoder \citep{kim_embeddistill_2023}. We hope that these findings will motivate the research community to reassess the trade-offs between method complexity and performance enhancements. Our code is publicly available in \href{https://github.com/Guest400123064/distill-retriever}{our GitHub repository}.

\section{The Trivially Simple Distillation Recipe}
  \subsection{Student Initialization}
  The initialization of student model weights is frequently not given enough attention in the knowledge distillation literature for IR. We find that a "well-prepared" student model can considerably alleviate distillation challenges. In this study, we investigate two classes of initialization approaches.

  \paragraph{Extract a Subset of Teacher Layers}
  In this initialization method, we establish the student model by taking a subset of the teacher model's transformer layers while keeping the embedding and pooling layers. By inheriting some of the teacher model's structural properties and knowledge, the student model is intuitively better prepared for efficient distillation in comparison to a randomly initialized student model. We conduct experiments using various combinations of teacher model layers to assess their impact on both performance and efficiency. Details would be discussed in \autoref{subsec:teacher-student}, \autoref{sec:results}, and \autoref{subsec:sublayer}.

  \paragraph{Adopt Other Pretrained Models}
  We also explore initializing the student model from other efficient pretrained models. Simultaneously, we investigate the influence of multiple factors, e.g., fine-tuning tasks and distance functions, on achieving a "well-prepared" initialization for the student model. We take DistilBERT \citep{distilbert} as our student model candidate and experiment with different DistilBERT models fine-tuned on diverse tasks. Examples include the \textcolor{blue}{\texttt{distilbert-base-uncased}} model and DistilBERT models fine-tuned on the MS MARCO dataset \citep{bajaj_ms_2018} with distinct objectives from \textcolor{blue}{\texttt{sentence-transformers}} on HuggingFace \citep{huggingface}. This approach enables us to evaluate the efficacy of using alternative pretrained models as starting points for student model initialization. Student model cards are listed in \autoref{subsec:mdl-ckpt}.
  
  \subsection{Embedding Alignment}
  DE-based IR systems often use vector similarity for searching \citep{DBLP:journals/corr/abs-1806-09823}, making it logical to match student and teacher embedding spaces.
  
  \paragraph{Contextualized Embedding Pooling Strategies} BERT-based encoders produce contextualized representations for all tokens from the input text. Common ways to aggregate token embeddings are selecting \texttt{[CLS]} embedding, computing mean values across all token embeddings, concatenating multiple pooled embeddings together, etc. We stick with the average pooling strategy for all presented experiment results in this paper, as in \citet{reimers_sentence-bert_2019}.

  \paragraph{Alignment Objective}
  Let $\text{Enc}_{\theta}^{s}(\cdot)$ denote the student query encoder parameterized by $\theta$ and $\text{Enc}^{t}(\cdot)$ denote the teacher query encoder, we minimize the expected Euclidean distance between the student and teacher embeddings,
  \begin{equation*}
      \mathcal{L}(\theta) = \mathbb{E}_{q \sim \mathcal{D}_q}\left[\lVert \text{Enc}_{\theta}^{s}(q) - \text{Enc}^{t}(q)\rVert_2\right]
  \end{equation*}
  Thus, $\theta$ is found by minimizing the empirical loss,
  \begin{equation*}
      \theta = \argmin_{\theta} \frac{1}{|\mathcal{Q}|} \sum_{q_i \in \mathcal{Q}} \lVert \text{Enc}_{\theta}^{s}(q_i) - \text{Enc}^{t}(q_i)\rVert_2
  \end{equation*}
  where $\mathcal{Q}$ denotes a set of queries sampled from the distillation domain. In our experiment, we set $\mathcal{Q}$ to be the queries of the IR datasets used by teacher query encoders. This simple optimization objective yields surprisingly performant student models \textbf{when paired with proper initialization.}

\section{Experiments}

  \subsection{Evaluation Datasets and Metrics}
  \paragraph{Retrieval Performance} For in-domain evaluation, we keep the dataset consistent with our teacher models' training corpus MS MARCO \cite{bajaj_ms_2018}. As for the out-of-domain (zero-shot) evaluation, we use the BEIR benchmark \citep{thakur_beir_2021} to evaluate our distillation method. It is a diverse collection of seven categories\footnote{The original publication presents nine categories, but the news and tweet retrieval datasets are not publicly available.} of IR-related tasks. We report normalized Discounted Cumulative Gain (nDCG@10) as the performance metric and average the \textbf{relative} performance drops to compare the distillation results.
  
  \paragraph{Inference Efficiency} We evaluate the efficiency of our distilled query encoder by measuring the wall-clock time required to process queries from the NQ dataset \citep{nq}. We simulate various scenarios, ranging from nearly online settings to batched processing, by selecting batch sizes of 4, 8, 16, 32, and 64. For each batch size, we record the elapsed time to process approximately $4 \times 10^3$ queries on a single Nvidia Tesla T4 GPU, repeating the process three times and taking the median time to calculate the number of queries processed per second as the evaluation result.
  
  \subsection{Teacher and Student Models}
  \label{subsec:teacher-student}
  \paragraph{The Teacher Model} In this work, we use a siamese DE model \textcolor{blue}{\texttt{msmarco-bert-base-dot-v5}} hosted on \href{https://huggingface.co/sentence-transformers/msmarco-bert-base-dot-v5}{the HuggingFace hub} for its competitive performance (\autoref{fig:teacher-performance}). The model was fine-tuned on MS MARCO using the dot score as the relevancy measurement and Margin Mean Squared Error (MarginMSE) as the objective function.

  \paragraph{Extractive Initialization} We select a total of thirteen combinations, comprising five combinations of 4-layer models, four combinations of 2-layer models, and four combinations of 1-layer models. The full combinations are listed in \autoref{subsec:sublayer}.
  
  \paragraph{DistilBERT Initialization} We explore six DistilBERT checkpoints. The students are initialized from the full model without extracting subsets of layers. Please refer to \autoref{subsec:mdl-ckpt} for the HuggingFace model cards. We discuss the potential relationship between distillation performance and model characteristics in \autoref{sec:results}.
  
  \subsection{Implementation Details}
  \label{subsec:impl-detail}
  We use the first 80\% of over eight million queries from the MS MARCO training set as our training data and the rest 20\% for validation. We train the student models using the AdamW optimizer \citep{adamw} for one epoch with Mean Squared Error (MSE) loss, applying a batch size of 128, a learning rate of $10^{-4}$ and $10^3$ warm-up steps.

\begin{table*}[htbp!]
    \centering
    \resizebox{\textwidth}{!}{
    \begin{tabular}{lcrcrcrcrcrcr}

\hline
Fine-tune Dataset        
    & \multicolumn{2}{c}{MS MARCO}        
    & \multicolumn{2}{c}{MS MARCO}        
    & \multicolumn{2}{c}{-}     
    & \multicolumn{2}{c}{MS MARCO}
    & \multicolumn{2}{c}{NLI + STS} 
    & \multicolumn{2}{c}{MS MARCO} \\

Fine-tune Objective     
    & \multicolumn{2}{c}{MarginMSE}       
    & \multicolumn{2}{c}{MarginMSE}       
    & \multicolumn{2}{c}{-} 
    & \multicolumn{2}{c}{MultiNegRanking} 
    & \multicolumn{2}{c}{CosineSimilarity} 
    & \multicolumn{2}{c}{MarginMSE} \\

Similarity Function 
    & \multicolumn{2}{c}{Dot} 
    & \multicolumn{2}{c}{Dot} 
    & \multicolumn{2}{c}{-}
    & \multicolumn{2}{c}{Cosine}
    & \multicolumn{2}{c}{Cosine}& \multicolumn{2}{c}{Cosine} \\
\hline \hline

\textbf{Dataset} $(\downarrow)$ \textbf{Ckpt}  $(\rightarrow)$ 
    & \multicolumn{2}{c}{\texttt{\textcolor{blue}{msmarco-dot}}}        
    & \multicolumn{2}{c}{\texttt{\textcolor{blue}{msmarco-tas-b}}}      
    & \multicolumn{2}{c}{\texttt{\textcolor{blue}{base-uncased}}} 
    & \multicolumn{2}{c}{\texttt{\textcolor{blue}{msmarco-base}}} 
    & \multicolumn{2}{c}{\texttt{\textcolor{blue}{nli-stb}}} 
    & \multicolumn{2}{c}{\texttt{\textcolor{blue}{msmarco-cos}}} \\
\hline \hline

MS MARCO (In-domain)
    & \textbf{.415} & \textbf{(-\ \ 6.58\%)} 
    & .412 & (-\ \ 7.17\%) 
    & .406 & (-\ \ 8.60\%) 
    & .389 & (-12.40\%) 
    & .389 & (-12.39\%) 
    & .346 & (-22.04\%) \\
\hline

TREC-COVID 
    & \textbf{.689} & \textbf{(-\ \ 7.58\%)} 
    & .676 & (-\ \ 9.30\%) 
    & .634 & (-14.86\%) 
    & .575 & (-22.78\%) 
    & .573 & (-23.03\%) 
    & .512 & (-31.31\%) \\

NFCorpus 
    & .290 & (-\ \ 7.84\%)
    & \textbf{.297} & \textbf{(-\ \ 5.87\%)} 
    & .283 & (-10.04\%) 
    & .271 & (-14.06\%) 
    & .269 & (-14.56\%) 
    & .230 & (-27.04\%) \\
\hline

NQ 
    & \textbf{.481} & \textbf{(-\ \ 7.94\%)} 
    & .480 & (-\ \ 8.04\%) 
    & .463 & (-11.42\%) 
    & .433 & (-17.09\%) 
    & .440 & (-15.75\%) 
    & .404 & (-22.67\%) \\

HotpotQA 
    & \textbf{.441} & \textbf{(-25.06\%)} 
    & .421 & (-28.38\%) 
    & .394 & (-33.06\%) 
    & .352 & (-40.12\%) 
    & .348 & (-40.87\%) 
    & .287 & (-51.15\%) \\

FiQA-2018 
    & \textbf{.291} & \textbf{(-\ \ 9.84\%)} 
    & .289 & (-10.61\%) & .282 & (-12.81\%) 
    & .269 & (-16.59\%) & .271 & (-16.15\%) 
    & .232 & (-28.18\%) \\
\hline

ArguAna 
    & \textbf{.426} & \textbf{(\ 14.33\%)} 
    & .417 & (\ 12.12\%) 
    & .392 & (\ \ \ 5.40\%) 
    & .408 & (\ \ \ 9.68\%) 
    & .368 & (-\ \ 1.07\%) 
    & .402 & (\ \ \ 8.09\%) \\

Touché-2020 
    & .232 & (-\ \ 2.22\%) 
    & .238 & (\ \ \ 0.59\%) 
    & .245 & (\ \ \ 3.36\%) 
    & \textbf{.247} & \textbf{(\ \ \ 4.22\%)} 
    & .244 & (\ \ \ 3.17\%) 
    & .234 & (-\ \ 1.43\%) \\
\hline

SCIDOCS 
    & \textbf{.121} & \textbf{(-17.02\%)} 
    & .119 & (-18.93\%) 
    & .110 & (-25.10\%) 
    & .099 & (-32.29\%) 
    & .100 & (-31.29\%) 
    & .086 & (-41.18\%) \\
\hline

CQADupStack 
    & \textbf{.218} & \textbf{(-15.54\%)} 
    & .212 & (-17.63\%) 
    & .204 & (-21.08\%) 
    & .180 & (-30.10\%) 
    & .185 & (-28.26\%) 
    & .148 & (-42.58\%) \\
    
Quora 
    & \textbf{.812} & \textbf{(-\ \ 3.59\%)} 
    & .809 & (-\ \ 3.98\%) 
    & .806 & (-\ \ 4.31\%)
    & .799 & (-\ \ 5.23\%) 
    & .792 & (-\ \ 6.04\%) 
    & .751 & (-10.88\%) \\
\hline

FEVER 
    & \textbf{.620} & \textbf{(-18.06\%)} 
    & .616 & (-18.64\%) 
    & .568 & (-25.00\%) 
    & .535 & (-29.39\%) 
    & .519 & (-31.40\%) 
    & .446 & (-41.05\%) \\

Climate-FEVER 
    & \textbf{.182} & \textbf{(-19.77\%)} 
    & .175 & (-22.64\%) 
    & .163 & (-28.12\%) 
    & .156 & (-30.89\%) 
    & .150 & (-33.70\%) 
    & .145 & (-35.77\%) \\

SciFact 
    & .541 & (-11.09\%) 
    & \textbf{.546} & \textbf{(-10.29\%)} 
    & .522 & (-14.23\%) 
    & .492 & (-19.14\%) 
    & .493 & (-19.02\%) 
    & .439 & (-27.76\%) \\

DBpedia-Entity 
    & \textbf{.337} & \textbf{(-12.41\%)} 
    & .329 & (-14.50\%) 
    & .314 & (-18.45\%) 
    & .287 & (-25.54\%) 
    & .302 & (-21.46\%) 
    & .265 & (-31.08\%) \\
\hline \hline

Avg. $\Delta$ Performance & \multicolumn{2}{c}{\textbf{-10.01\%}}  &
\multicolumn{2}{c}{-10.88\% } & 
\multicolumn{2}{c}{-14.55\% } & 
\multicolumn{2}{c}{-18.78\% } & 
\multicolumn{2}{c}{-19.45\% } & 
\multicolumn{2}{c}{-27.07\% } \\

    \hline
    \end{tabular}%
}
    \caption{The DistilBERT-based students' nDCG@10 and percentage change compared to the teacher model across BEIR evaluation datasets. The models are ordered (left to right) according to their average performance degradation.}
    \label{tab:distilbert_perf}
  \end{table*}
  
\section{Results and Discussions}
  \label{sec:results}
  \paragraph{Initializing from Subsets of Teacher Layers} \autoref{fig:extractive-performance} illustrates the performance of the distilled query encoders. We observe that different initialization strategies can lead to up to 6\% variability in performance, even with the same number of layers. However, \textbf{we find that initializing the students with the first and last few layers consistently yields preferable results}, which aligns with previous findings \citep{fan2019reducing, Sajjad2020PoorMB, dong-etal-2022-exploring}. For instance, considering the 1-layer student encoder (\autoref{fig:1L-performance}), initializing from the last layer yields the best outcomes across all datasets except for ArguAna \citep{arguana} and Touché-2020 \citep{touche}, preserving an average relative performance of 86.1\%. This observation applies similarly to the 2-layer (retaining the first and last layers) and 4-layer (retaining the first and last two layers) students, which exhibit performance preservation rates of 92.5\% and 96.2\% respectively, aligning closely with the performance of the supervised distilled 6-layer encoder at 96.6\% \citep{kim_embeddistill_2023}.

  \paragraph{Initializing from DistilBERTs} The results in \autoref{tab:distilbert_perf} reveal the within-group performance comparison. Since all student models undergo the same embedding-alignment distillation process, the final performance preservation rate can serve as a proxy for the "well-preparedness" of students. \texttt{\textcolor{blue}{msmarco-dot}} performs the best. Its tuning configuration is the same as its teacher's, i.e., the same dataset, distance function, and objective function. \texttt{\textcolor{blue}{msmarco-tas-b}}, tuned with the balanced topic-aware sampling technique \citep{hofstatter_efficiently_2021}, closely follows. Such a variation poses a slightly greater challenge in embedding alignment. On the other side of the spectrum, changing a distance measurement alone makes alignment drastically harder, as shown from \texttt{\textcolor{blue}{msmarco-cos}}. Interestingly, using a different objective function (\texttt{\textcolor{blue}{msmarco-base}} and \texttt{\textcolor{blue}{nli-stsb}}) appears to alleviate misalignment, suggesting the potential interaction between objective and distance functions. Additionally, a clean, pretrained-only student (\texttt{\textcolor{blue}{base-uncased}}) performs better when a perfect replication of the teacher's fine-tuning setting is not present. Notably, all DistilBERT-based students perform worse than the top-performing extractive students. The 2-layer extractive student outperforms the 6-layer \texttt{\textcolor{blue}{msmarco-dot}} with a performance gap of 2.5\%.

  \paragraph{Where are the Well-prepared Students?} Student pretraining has been demonstrated to be crucial for knowledge distillation in language understanding tasks \citep{well-read-student}. However, in our asymmetric DE system, the student encoder operates in conjunction with the document encoder of the teacher system, deviating from the conventional distillation procedure. In this case, \textbf{the effectiveness of student models lies not in their sheer capability but rather in their compatibility.} \citet{dong-etal-2022-exploring} employed t-SNE \citep{tsne} to visualize the embedding spaces of DE encoders in the context of QA tasks. They observed that the two encoders of an asymmetric system tend to map questions and answers onto distinct parameter spaces, even when trained jointly. This observation elucidates the reason why extractive initialization significantly reduces the difficulty of knowledge distillation in our scenario. Furthermore, we extend these findings by demonstrating that aligning the training objectives, similarity measures, and fine-tuning datasets with those of the teacher model can enhance embedding space compatibility. Note that fine-tuning on similar tasks without aligning other elements, e.g., the distance function, may undermine compatibility. Our findings, in conjunction with the results from \citet{kim_embeddistill_2023}, suggest that supervision signals play a crucial role in alignment while parameter-sharing inherently addresses this issue.

\begin{figure}[htp!]
    \centering
    \includegraphics[width=\linewidth]{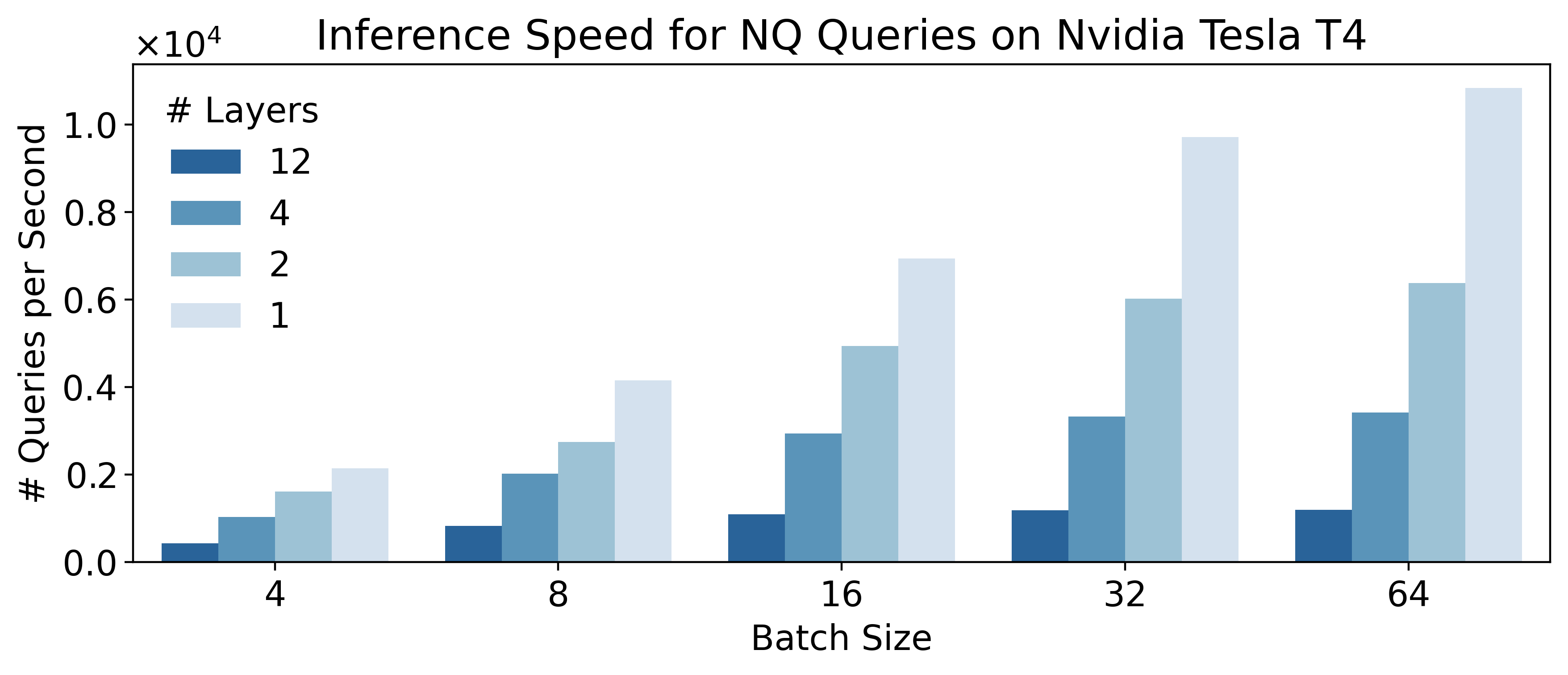}
    \caption{Inference speeds of the distilled query encoders compared to that of the full teacher model. The improvements in inference efficiencies become more drastic as batch size increases.}
    \label{fig:inference-speed}
  \end{figure}

  \paragraph{Inference Efficiency} \autoref{fig:inference-speed} shows that student models initialized from a subset of teacher layers have significantly improved inference speed compared to the teacher model, even with small batch sizes. Considering the marginal performance loss, query encoder distillation provides substantial benefits over the siamese DE encoder.
  
\section{Related Work}
  \paragraph{Efficient Methods for DE-based IR Systems} Various techniques have been proposed to enhance encoder performance in IR systems, including knowledge distillation \citep{hofstatter_improving_2021, zeng_curriculum_2022, lin_prod_2023, kim_embeddistill_2023}, improved pretraining objectives \citep{DBLP:journals/corr/abs-1906-00300, DBLP:journals/corr/abs-2002-03932, gao_unsupervised_2021, DBLP:journals/corr/abs-2112-09118}, data augmentation \citep{DBLP:journals/corr/abs-2107-13602}, better sampling techniques \citep{lin_-batch_2021, zhang_adversarial_2022}, ensembles \citep{hofstatter_efficiently_2021, lin_how_2023, ren_rocketqav2_2021}. However, most of these methods focus on siamese architectures, as asymmetric DE pairs are prone to representation collapse \citep{Leonhardt2022DistributionAlignedFF} or misalignment of embedding spaces \citep{dong-etal-2022-exploring}, making them challenging to train. Due to the shared parameters between query and document encoders, practitioners often need to constrain model size for practicality in production settings, despite the significance of larger models for better retrieval and generalization performance \citep{ni-etal-2022-large, yu2022coco}. Consequently, this constraint often leads to complex training procedures. In contrast, our simple recipe adopts the train-large-distill-small paradigm, offering a straightforward and effective approach to model development and can be adopted out of the box for existing systems.

  \paragraph{Embedding Alignment for IR} Concurrently, \citet{kim_embeddistill_2023} propose incorporating embedding alignment loss into the supervised distillation pipeline. However, they initialized models from other checkpoints without recognizing the importance of using teacher weights as initialization. Additionally, \citet{campos_quick_2023} suggest minimizing the KL divergence between student and teacher embeddings in an unsupervised manner. Yet, to the best of our knowledge, they do not explore the impact of different layer subsets, whereas our work demonstrates the significant variance caused by such choices.

\section{Conclusion}
  In this work, we leverage the characteristics of typical production DE-based IR systems to propose a minimalistic baseline method for improving online efficiency through embedding-alignment distillation. We explore the significance of student initialization for asymmetric DEs and demonstrate that a "well-prepared" student can achieve over five times improvement in efficiency with only 7.5\% average performance degradation. We also observe that "well-prepared" students generally have aligned embedding spaces with their teachers, and a simple approach to construct such students is by extracting the first and last few layers from the teacher models. Our findings aim to enhance the accessibility of neural IR systems and encourage the research community to reassess the trade-offs between method complexity and performance improvements.

\section*{Limitations}
  \label{sec:limitation}
  \paragraph{Limited Experimental Scope} Our study's experimental scope was limited to testing distilled student models against a single teacher model. A more comprehensive evaluation would involve multiple teacher models of varying sizes, fine-tuning tasks, and datasets. Additionally, in our experiment with DistilBERT-based student models, incorporating more checkpoints would enable a more thorough comparison across different factors.

  \paragraph{Unexplored Embedding Size Variations} We kept the embedding size (768) consistent across student models to maintain variable consistency. Future research could investigate student models with different embedding sizes to determine if the observed trends hold true across models of varying widths.
  
  \paragraph{Lack of Error Analysis} A common distillation limitation, as noted by \citet{hooker2020characterising}, is the considerable performance decline for certain data subsets. In our study, we couldn't conduct a thorough error analysis due to the lack of appropriate tools for comparing individual data points in retrieval tasks.

\section*{Ethics Statement}
Although our method improves accessibility for IR systems, it is essential to evaluate whether the proposed approach might introduce biases or unfairness in the retrieval results. As our work lacks extensive error analysis, we cannot entirely rule out the possibility that distilled query encoders may discard certain hard-to-process cases critical for ensuring fairness across various query topics and user groups. A comprehensive error analysis would be beneficial in future research to identify and address potential biases in the distilled query encoders, ultimately fostering fair and unbiased retrieval results for all users.

\section*{Acknowledgements}
We thank the anonymous reviewers for their time and constructive feedback. Also, we would like to thank Prof. Mark Yatskar for his precious suggestions during the development of this work.

\newpage
\bibliography{custom}
\bibliographystyle{acl_natbib}

\appendix

\section{Technical Details}
  \begin{figure*}[bp!]
    \centering
    \includegraphics[width=\textwidth]{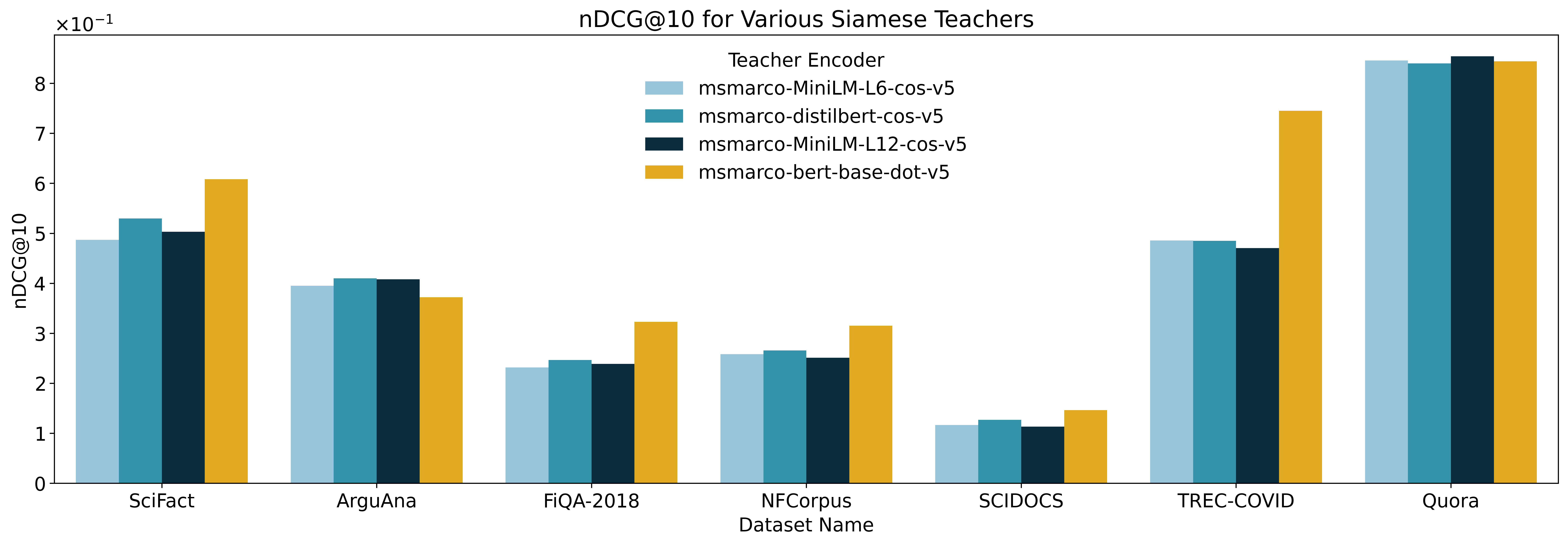}
    \caption{The performances of various teachers provided by SentenceTransformers on a subset of BEIR benchmarks. We select the teacher with the highest retrieval performance.}
    \label{fig:teacher-performance}
  \end{figure*}

  \subsection{Layer-subtraction Schemes}
  \label{subsec:sublayer}
  The 4-layer models were initialized using the following schemes: [1,4,7,10], [0,1,10,11], [0,1,2,3], [4,5,6,7], and [8,9,10,11]. The first two schemes were inspired by the results from \citet{fan2019reducing}, which suggested that the input and output layers are often more influential in embedding representations than the middle layers. The latter three schemes were used to validate this intuition and guide our selection schemes for 2-layer and 1-layer initialization. Combinations of 2-layer include [0, 10], [0, 11], [1, 10], and [1, 11]. Layers extracted to make 1-layer models are [0], [1], [10], and [11].
  
  \subsection{DistilBERT-based Student Checkpoints}
  \label{subsec:mdl-ckpt}
  The HuggingFace model cards of the DistilBERT checkpoints adopted in the experiments are listed below. Except for \textcolor{blue}{\texttt{distilbert-base-uncased}}, all other models have \textcolor{blue}{\texttt{sentence-transformers/}} prefix. The same order also maps to \autoref{tab:distilbert_perf}.
  \begin{enumerate}[noitemsep]
      \item \textcolor{blue}{\texttt{msmarco-distilbert-dot-v5}}
      \item \textcolor{blue}{\texttt{msmarco-distilbert-base-tas-b}}
      \item \textcolor{blue}{\texttt{distilbert-base-uncased}}
      \item \textcolor{blue}{\texttt{msmarco-distilbert-base-v3}}
      \item \textcolor{blue}{\texttt{distilbert-base-nli-stsb-mean-tokens}}
      \item \textcolor{blue}{\texttt{msmarco-distilbert-cos-v5}}
  \end{enumerate}

  \begin{figure*}[bp!]
    \centering
    \includegraphics[width=\textwidth]{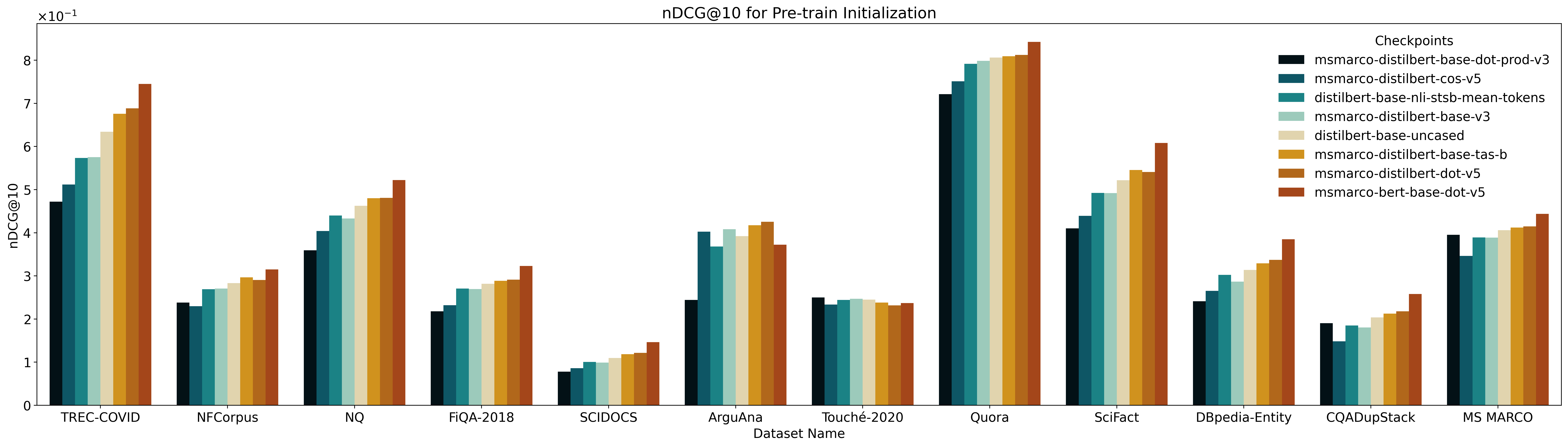}
    \caption{In general, students initialized from pretrained models perform worse than direct layer extraction.}
    \label{fig:pretrain-performance}
  \end{figure*}

\section{Additional Results}
  \subsection{Other Visualizations}
  \begin{figure*}[t!]
    \centering
    \includegraphics[width=\textwidth]{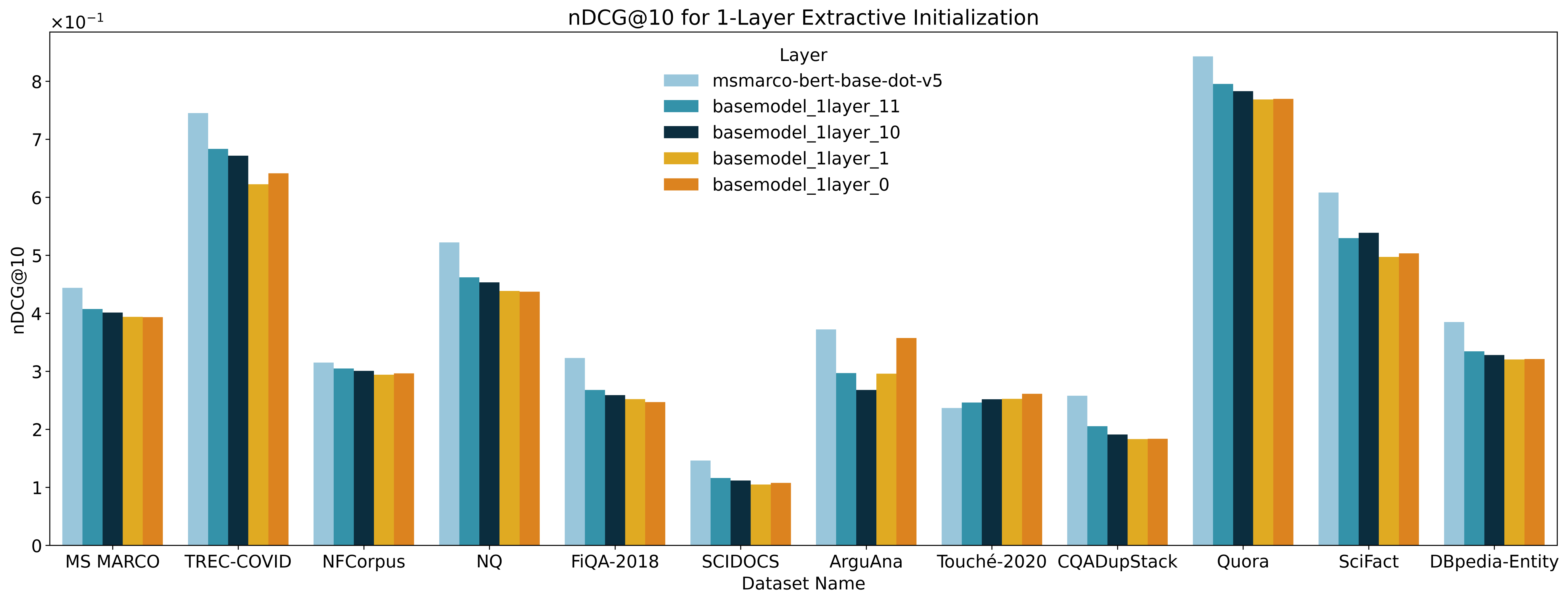}
    \caption{Deeper layers are more preferable than the shallower layers in terms of initialization strategy.}
    \label{fig:1L-performance}
  \end{figure*}
  
  \begin{figure*}[t!]
    \centering
    \includegraphics[width=\textwidth]{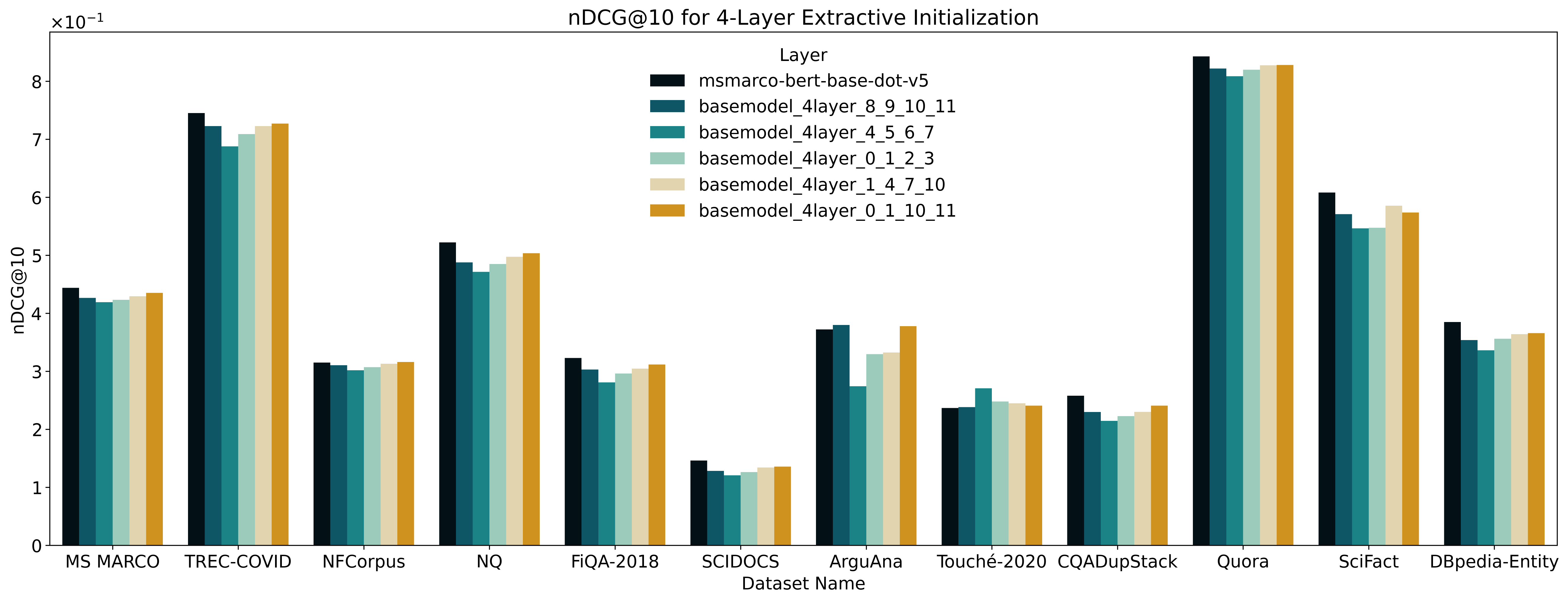}
    \caption{The first few layers and last few layers are more preferable in terms of initialization strategy.}
    \label{fig:4L-performance}
  \end{figure*}
  
  \autoref{fig:teacher-performance} shows the performances of various teachers provided by SentenceTransformers on a subset of BEIR benchmarks. We select the teacher with the highest retrieval performance \texttt{\textcolor{blue}{msmarco-bert-base-dot-v5}}. \autoref{fig:pretrain-performance} shows the performances of the student models initialized from other DistilBERT checkpoints. In general, students initialized from pretrained models perform worse than direct layer extraction. \autoref{fig:1L-performance} and \autoref{fig:4L-performance} demonstrate that The first few layers and last few layers are more preferable in terms of initialization strategy.

\end{document}